\documentclass[preprint]{elsarticle}
\usepackage{amssymb}
\usepackage{amsthm}

\usepackage{lineno}

\journal{PLA}

\begin{document}

\begin{frontmatter}

\title{Interplay of Electron-Electron and Electron-Phonon Interactions in Molecular Junctions   }

\author[sbu]{Amir Eskandari-asl\corref{cor1}}
\ead{a{\_}eskandariasl@sbu.ac.ir; amir.eskandari.asl@gmail.com}
\address[sbu]{Department of physics, Shahid Beheshti University, G. C. Evin, Tehran 1983963113, Iran}

\begin{abstract}
In this work we consider a current carrying molecular junction with both electron-phonon and electron-electron interactions taken into account. After performing Lang-Firsov transformation and considering Markov approximations in accordance to weak coupling to the electronic leads, we obtain the master equation governing the time evolution of the reduced density matrix of the junction. The steady state of the density matrix can be used to obtain I-V characteristic of the junction in several regimes of strengths of the interactions. Our results indicate that the system can show negative differential conductance (that is, the current decreases by increasing the applied bias voltage) in some regimes as an interplay between the electron-phonon and Coulomb interactions. 
\end{abstract}

%\keywords{master equation; negative differential conductance; Coulomb blockade}%Use showkeys class option if keyword
                              
\end{frontmatter}

%\tableofcontents
\section{Introduction}
In recent years, the technology made it possible to miniaturize the size of quantum dots(QD) to molecular scales and investigate the so called molecular junctions (MJ)\cite{cuniberti,cuevas2017}. These small junctions have attracted many attentions both in experimental and theoretical sides\cite{terrones2002,zhitenev2002,nitzan2003,galperin2006,tao2006,galperin2008,hartle2011}. 

MJs have several interesting transport properties, such as bi-stability, negative differential conductance (NDC), etc\cite{CLi1,Elor,Lilj,le2003negative}. NDC is of special importance, because of its potential to be used in designing future molecular sized circuits. On the other hand, from the point of view of fundamental physics, understanding such phenomena is of great value. Indeed, several experimental setups are designed to investigate NDC\cite{xu2015negative}. 

One of the main characteristics of MJs in comparison to large QDs is the strong coupling of electrons to the vibrations of molecule. This electron-phonon (e-ph) interaction is one of the most important candidates to theoretically explain NDC\cite{galperin2005,galperin2008non,eskandari2016bi}. On the other hand, Coulomb interaction can also be the origin of NDC in some systems where the LUMO and HOMO levels have different couplings to the leads\cite{muralidharan2007generic}. It should be noted that because of the small size of MJs, the Coulomb repulsion is usually very strong and it is justified to ignore double occupancy of the levels. However, in some cases e-e and e-ph interactions are of comparable strengths , which require theoretical considerations\cite{ren2012thermoelectric,de2016}.

In MJs with weak coupling to the electronic leads, one can trace out the degrees of freedom of the leads and obtain a master equation(ME) which describes the dynamics of MJ (electrons and phonons)\cite{hartle2011,schaller2013,sowa2017,kosov2017,eskandari2018influence}. This approach is extendable to the case where we have both e-e and e-ph interactions\cite{de2016}. In this work, we consider such a problem, and after Lang-Firsov(LF) transformation, we obtain the ME which governs the time evolution of our MJ. It is seen that in steady state, the interplay between e-e and e-ph interaction results in NDC at some bias voltages.

The paper is organized as follow. In Sec.\ref{mm}, the Hamiltonian of our system is introduced and the corresponding ME is derived. Moreover, the formula to obtain the electrical current through the MJ is obtained. In Sec.\ref{nr}, we present our numerical results and discussions, and finally, Sec.\ref{con} concludes our work.   

\section{Model and Method} \label{mm}
Our model consists of a single level MJ which connects two leads. This level can be populated by two opposite spin electrons which repel each other according to Coulomb interaction. Moreover, electrons on the MJ are coupled to a single frequency phonon mode. The Hamiltonian of this system is
\begin{eqnarray}
&&\hat{H}=\hat{H}_{d}+\hat{H}_{leads}+\hat{H}_{ph}+\hat{H}_{tun}+\hat{H}_{e-ph},
\label{hamt}
\end{eqnarray}
\begin{eqnarray}
&&\hat{H}_{d}=\sum_{\sigma=\uparrow , \downarrow}\epsilon_{0} \hat{n}_{\sigma}+ U_{0} \hat{n}_{\uparrow} \hat{n}_{\downarrow},
\label{hd}
\end{eqnarray}
\begin{eqnarray}
&&\hat{H}_{leads}=\sum_{k,\sigma,\alpha \in \left\lbrace R,L\right\rbrace } \epsilon_{k,\alpha} \hat{a}^{\dag}_{k\sigma \alpha} \hat{a}_{k \sigma \alpha} ,
\label{hleads}
\end{eqnarray}
\begin{eqnarray}
&&\hat{H}_{ph}=\Omega \hat{b}^{\dag}\hat{b},
\label{hph}
\end{eqnarray}
\begin{eqnarray}
&&\hat{H}_{tun}=\sum_{k,\sigma,\alpha \in \left\lbrace R,L\right\rbrace } V_{k\alpha} \hat{c}_{\sigma}^{\dag} \hat{a}_{k\sigma\alpha}+ h.c. ,
\label{htun}
\end{eqnarray}
\begin{eqnarray}
&&\hat{H}_{e-ph}=\sum_{\sigma} \lambda \Omega \hat{n}_{\sigma} \left(\hat{b}+ \hat{b}^{\dag} \right),
\label{heph}
\end{eqnarray}
where $ \hat{c}_{\sigma} $ ($ \hat{c}^{\dag}_{\sigma}  $) is the annihilation (creation) operator of an electron with spin $ \sigma $ on MJ, $ \hat{n}_{\sigma}=\hat{c}^{\dag}_{\sigma}\hat{c}_{\sigma} $ is the number operator, $ \epsilon_{0} $ is the onsite energy of electrons on the MJ and $ U_{0} $ is the strength of Coulomb repulsion. $ \hat{b} $($ \hat{b}^{\dag} $) is the annihilation (creation) operator of phonons on MJ, $ \Omega $ is the phonon frequency and $ \lambda $ determines electron-phonon coupling. Moreover, $ \hat{a}_{k\sigma\alpha} $ ($ \hat{a}^{\dag}_{k\sigma\alpha} $) annihilates (creates) an electron with spin $ \sigma $ in the state $ k $ of the lead $ \alpha $ ($ \alpha=R,L $), and  $ V_{k,\alpha} $ determines the electron hopping between MJ and the leads.

We perform the LF transformation on our Hamiltonian (see the appendix) as a result of which the e-ph part disappears. The transformed Hamiltonian of the system is
\begin{eqnarray}
&&\hat{\tilde{H}}=\sum_{\sigma=\uparrow , \downarrow}\epsilon \hat{n}_{\sigma}+ U \hat{n}_{\uparrow} \hat{n}_{\downarrow}+\Omega \hat{b}^{\dag}\hat{b} \quad\nonumber\\
&&+\sum_{k,\sigma,\alpha \in \left\lbrace R,L\right\rbrace } \left(  V_{k\alpha}  \hat{c}_{\sigma}^{\dag} \hat{X} ^{\dag} \hat{a}_{k\sigma\alpha}+ h.c.\right) + \hat{H}_{leads},
\label{lfh}
\end{eqnarray}
where $\epsilon=\epsilon_{0}-\lambda^{2} \Omega \nonumber$ and $U=U_{0}-2 \lambda^{2} \Omega $.

Following the standard steps for deriving a Markovian ME in the limit of weak lead to MJ coupling, we can obtain dynamics of the density matrix (DM) of the MJ,$ \rho $ (including both electrons and phonons). In order to do such calculations, we work in the many-body Fock space, spanned by the number states of the form $ \vert n_{\uparrow},n_{\downarrow},m \rangle $, where $ n_{\sigma}=0,1 $ is the electron number and $ m=0,1,2,... $ is the number of phonons(in numerical computations one has to consider an upper bound for this number of phonons). As in our former work\cite{eskandari2018influence}, it is straightforward to show that if the initial DM is diagonal, it will remain diagonal for all times. A diagonal DM corresponds to a mixture(and not superposition) of number states. For example, states with definite number of electrons and phonons, including the empty MJ, correspond to diagonal DMs. It is noteworthy that these number states are pointer states\cite{zurek} of our system under the interaction with the leads, so we expect the final steady state to be a mixture of number states(which corresponds to a diagonal DM) even for an initially non-diagonal DM. Therefore, it suffices to just obtain the time evolution of the diagonal elements of DM, $\langle n_{\uparrow},n_{\downarrow},m \vert \rho \vert n_{\uparrow},n_{\downarrow},m \rangle  $, which we show by $ P_{n_{\uparrow},n_{\downarrow},m} $.  The ME of the system can be shown to be 
\begin{eqnarray}
&&\frac{d}{dt} P_{00m}=\sum_{m^{\prime},\alpha } \Gamma_{\alpha} \left( \left[ 1-f_{\alpha}\left( \Omega \left( m^{\prime}-m\right)  +\epsilon\right)\right] \vert \hat{X}_{mm^{\prime}} \vert^{2} P_{10m^{\prime}}\right.\quad\nonumber\\
&& \left.+ \left[ 1-f_{\alpha}\left( \Omega \left( m^{\prime}-m\right)  +\epsilon\right)\right] \vert \hat{X}_{mm^{\prime}} \vert^{2} P_{01m^{\prime}}\right.\quad\nonumber\\
&&\left.- 2 f_{\alpha}\left( \Omega \left( m^{\prime}-m\right)  +\epsilon\right) \vert \hat{X}_{mm^{\prime}} \vert^{2} P_{00m} \right) ,
\label{dp00m}
\end{eqnarray}   

\begin{eqnarray}
&&\frac{d}{dt} P_{10m}=\sum_{m^{\prime},\alpha } \Gamma_{\alpha}  \left( f_{\alpha}\left( \Omega \left(m- m^{\prime}\right)  +\epsilon\right) \vert \hat{X}_{m^{\prime}m} \vert^{2} P_{00m^{\prime}}\right.\quad\nonumber\\
&& \left.+\left[ 1-f_{\alpha}\left( \Omega \left( m^{\prime}-m\right)+ U  +\epsilon\right)\right] \vert \hat{X}_{mm^{\prime}} \vert^{2} P_{11m^{\prime}}\right.\quad\nonumber\\
&& \left.-\left[ 1-f_{\alpha}\left( \Omega \left(m-m^{\prime}\right)  +\epsilon\right)\right] \vert \hat{X}_{m^{\prime}m} \vert^{2} P_{10m}\right.\quad\nonumber\\
&&\left.- f_{\alpha}\left( \Omega \left( m^{\prime}-m\right)+ U +\epsilon\right) \vert \hat{X}_{mm^{\prime}} \vert^{2} P_{10m} \right) , 
\label{dp10m}
\end{eqnarray}   

\begin{eqnarray}
&&\frac{d}{dt} P_{11m}=\sum_{m^{\prime},\alpha } \Gamma_{\alpha}  \left( f_{\alpha}\left( \Omega \left(m- m^{\prime}\right)+ U  +\epsilon\right)\right.\quad\nonumber\\
&& \left. \vert \hat{X}_{m^{\prime}m} \vert^{2} \left( P_{10m^{\prime}}+P_{01m^{\prime}}\right) \right.\quad\nonumber\\
&& \left.- 2 \left[ 1-f_{\alpha}\left( \Omega \left(m-m^{\prime}\right)+ U +\epsilon\right)\right] \vert \hat{X}_{m^{\prime}m} \vert^{2} P_{11m} \right) ,
\label{dp11m}
\end{eqnarray}  
 
where $ f_{\alpha}(\omega)=\frac{1}{e^{\beta_{\alpha}(\omega-\mu_{\alpha})}+1} $ is the Fermi distribution of lead $ \alpha $, in which $ \mu_{\alpha} $ is chemical potential of the lead and $ \beta_{\alpha} $ is its inverse temperature. $ \Gamma_{\alpha} $ determines the tunneling rate of electrons between MJ and lead $ \alpha $, which is defined to be $ \Gamma_{\alpha}(\omega)=\sum_{k} 2 \pi \vert V_{k\alpha} \vert^{2} \delta(\epsilon_{k\alpha}-\omega) $. In wide band limit(WBL), we take $ \Gamma_{\alpha} $ to be independent of $ \omega $. Moreover, it should be noticed that according to spin symmetry in our model, $ P_{10m}=P_{01m} $. As it is defined in the appendix, $ \hat{X}\equiv\exp [\lambda(\hat{b}-\hat{b}^{\dag})]  $, and $ \hat{X}_{m m^{\prime}} $ is its  matrix element in the basis of phonon number states, i.e., $ \hat{X}_{m m^{\prime}}=\langle m\vert \hat{X} \vert m^{\prime}\rangle $ . By expanding the exponential and doing the algebra, one can show that
\begin{eqnarray}
&&\vert \hat{X}_{m m^{\prime}} \vert^{2}= e^{-\lambda^{2}} \vert\sum_{j=0}^{m_{<}} \frac{(-1)^{j}\lambda^{2j+m_{>}-m_{<}} \sqrt{m!m^{\prime}!}}{j!(j+m_{>}-m_{<})!(m_{<}-j)!}\vert^{2},
\label{khi}
\end{eqnarray}
where $ m_{<} $ and $ m_{>} $ are the minimum and maximum of $ m $ and $ m^{\prime} $, respectively.

The diagonal element $ P_{n_{\uparrow},n_{\downarrow},m} $ determines the probability of having the MJ filled with $ m $ phonons and $ n_{\uparrow} $ and $ n_{\downarrow} $ electrons with spins $ \uparrow $ and $ \downarrow $, respectively. The Eqs.\ref{dp00m}-\ref{dp11m}, describe time evolution of these probabilities by tunneling of electrons into and out of MJ. Suppose that the state of MJ at some time is $\vert n_{\uparrow},n_{\downarrow},m \rangle$ with energy $ N_{e}\epsilon+ m \Omega+\delta_{N_{e},2} U $, and by a random hopping of one electron it changes into $ \vert n_{\uparrow}^{\prime},n_{\downarrow}^{\prime},m^{\prime} \rangle $ with energy $ N_{e}^{\prime}\epsilon+ m^{\prime} \Omega+\delta_{N_{e}^{\prime},2}U $  (here, $ N_{e}=n_{\uparrow}+n_{\downarrow} $). The difference of the energy of these two states, should be compensated by the leads.  For an electron to tunnel from MJ to the lead $ \alpha $, a level in the lead with appropriate energy has to be empty, and this is the origin of the coefficients of the form $1-f_{\alpha}(\omega)$. For the reverse direction, i.e., tunneling of the electrons into the MJ, the corresponding level of the lead should be full, so the terms of the form $ f_{\alpha}(\omega)$ can be understood. 

In Eq.\ref{dp00m}, the rate of change of the probability of having no electrons and $ m $ phonons in MJ is given. As it is suggested by the first two terms on the right side, this probability is increased by the processes of tunneling of one electron with either spin from the singly occupied MJ to an energy level in the leads that compromises with the initial and final number of phonons. On the other hand, the negative terms show that the reduction of this probability is described by the processes in which one electron hops into the MJ with initially no electrons. Similar physical explanations can be given for Eqs.\ref{dp10m} and \ref{dp11m}. It is noteworthy that in the limit of weak coupling between the MJ and leads, we just consider one electron processes.

The number of electrons in MJ is $ N_{e}=\sum_{m} \left( P_{10m}+ P_{01m}+2 P_{11m} \right)  $. The electrical currents from the leads to the MJ determine the rate of change of electron population, i.e., $ dN_{e}/dt=\sum_{\alpha} I_{\alpha} $. Comparing with Eqs.\ref{dp00m}-\ref{dp11m}, the relation for current from lead $ \alpha $ to the MJ is obtained as
\begin{eqnarray}
I_{\alpha}&=& 2 \Gamma_{\alpha}\sum_{mm^{\prime}} \left( f_{\alpha}\left( \Omega \left(m^{\prime}-m\right)  +\epsilon\right) \vert \hat{X}_{mm^{\prime}} \vert^{2} P_{00m}\right.\quad\nonumber\\
&& \left.-\left[ 1-f_{\alpha}\left( \Omega \left(m-m^{\prime}\right)  +\epsilon\right)\right] \vert \hat{X}_{m^{\prime}m} \vert^{2} P_{10m}\right.\quad\nonumber\\
&&\left.+ f_{\alpha}\left( \Omega \left( m^{\prime}-m\right)+ U +\epsilon\right) \vert \hat{X}_{mm^{\prime}} \vert^{2} P_{10m} \right.\nonumber\\
&& \left.-  \left[ 1-f_{\alpha}\left( \Omega \left(m-m^{\prime}\right)+ U +\epsilon\right)\right] \vert \hat{X}_{m^{\prime}m} \vert^{2} P_{11m} \right),\qquad 
\label{ecurr}
\end{eqnarray}
and the total current passing through the MJ is $ I=(I_{L}-I_{R})/2 $. 

In the next section, we numerically compute the steady-state current as a function of the applied bias voltage and investigate the situation when e-e and e-ph interactions coexist.

\section{Numerical Results} \label{nr}
In this section we represent our numerical results for the steady state currents. We consider an initially empty MJ at each bias voltage and numerically compute the time evolution of the diagonal elements until we  achieve the steady state. For our numerical calculations we consider the maximum phonon number to be 60, and define the steady state as a time at which the greatest rate of change of the diagonal elements is of the order of $ 10^{-5} $. It should be noted that we tested the reliability of our results by changing both of these numbers for some bias voltages and checking that the results won't change noticeably. We work in a system of units in which $ e=\hbar=1 $. Also, the Boltzmann constant, $ k_{B} $, is taken to be 1. We set the phonon frequency to be our energy unit, i.e., $ \Omega=1 $. These automatically set our units of time, bias voltage and electrical current. Moreover, the bias voltage is applied symmetrically, so that $ \mu_{L}=-\mu_{R}=V/2 $, and we consider $ \epsilon=-0.25 $. In this work, we don't consider the temperature gradient between the leads and set both leads to be at zero temperature. Finally, the tunneling rates between MJ and the leads are assumed to be $ \Gamma_{L}=\Gamma_{R}=0.1 $. 

\begin{figure}  %[ht!]
\includegraphics{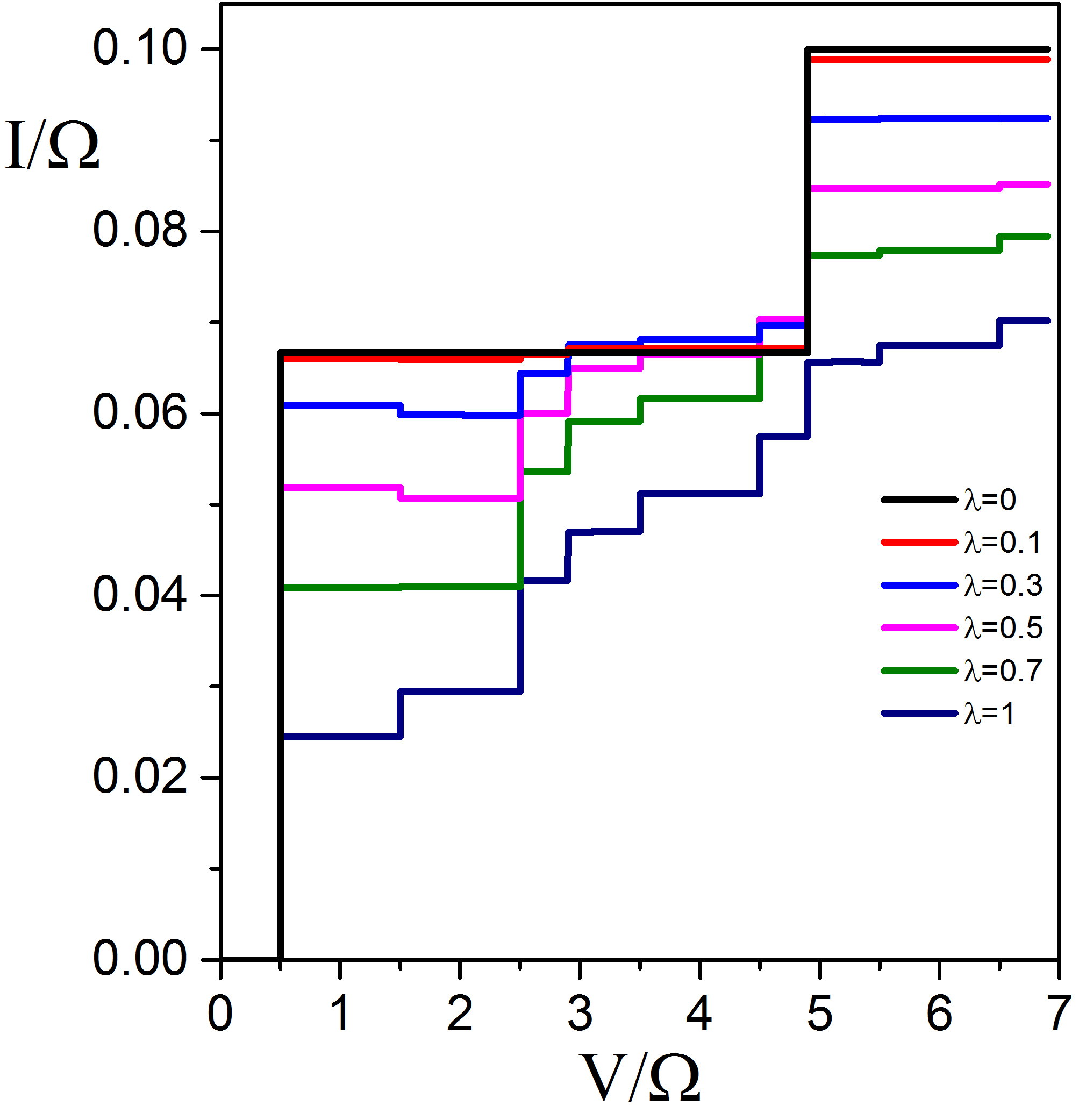}
\caption{\label{fig1} The current as a function of bias voltage, for different values of $ \lambda $, while $ U=2.7 $ and $ \epsilon=-0.25 $. By increasing the e-ph coupling strength, NDC at first gets stronger at the bias voltage of $ V=2 (\epsilon+\Omega) $, and then gradually disappears. }
\end{figure}

In Fig.\ref{fig1}, the current through MJ as a function of bias voltage is depicted for different values of e-ph coupling, $ \lambda $, ranging from 0 to 1. In all of these curves, the electrical current starts when the chemical potential of the right lead reaches $ \epsilon $, that is, when the bias voltage is $ V=0.5 $. For the case where $ \lambda $ vanishes, there is just one more step in the I-V curve, and that is when $ V/2=\epsilon+U $, as is expected in Coulomb blockade regime.

When $ \lambda $ is not zero, all of the steps in the electrical current occur at the side-band energies, that is, every step is at a bias voltage of the form $ \vert \frac{V}{2}\vert = \epsilon+ m \Omega+ n U $, where $m$ is integer and $ n=0,1 $. The most interesting ones of these steps, are those that show NDC, which in Fig.\ref{fig1} are at the bias voltage of $ V=2 (\epsilon+\Omega) $. 

As can be seen from Fig.\ref{fig1}, by increasing the e-ph coupling strength, this NDC increases at first, and then gradually disappears. In order to understand this behavior, we note that before this bias voltage, non of the phonon side-bands lie in the bias window. At $ V=2 (\epsilon+\Omega) $, another transport channel is opened and phonons get excited. As a result, we have two opposing phenomena. Since a new transport channel is opened the current tends to increase. On the other hand, the tunneling electron couples to the phonons and spends more time in the junction (this results in the increasing of electron population on the MJ, as can also be seen in a spin-less model, like what we used in our former work\cite{eskandari2018influence}). Because of Coulomb interaction, the increased population of electrons with one spin blocks opposite spin electrons to pass through, which reduces the current (it should be noticed that at this bias voltage, double occupancy is prohibited). The relative strength of e-e and e-ph interactions determines which effect would win. For small $ \lambda $s, the transport channel is of less importance than the Coulomb repulsion, however, when the $ \lambda $ gets strong enough, the new opened channel becomes more important. Moreover, by increasing e-ph coupling the current before the step also reduces. This current which is for bias voltages $ \epsilon < V/2 < \epsilon+\Omega $, can be computed analytically to be (in this voltage range no phonons are exited and $ P_{n_{\uparrow},n_{\downarrow},m} $ vanishes unless $ m=0 $)   
\begin{eqnarray}
I_{0}=\frac{2}{3} \Gamma e^{-\lambda^{2}}.
\label{i0}
\end{eqnarray}

\begin{figure}  %[ht!]
\includegraphics{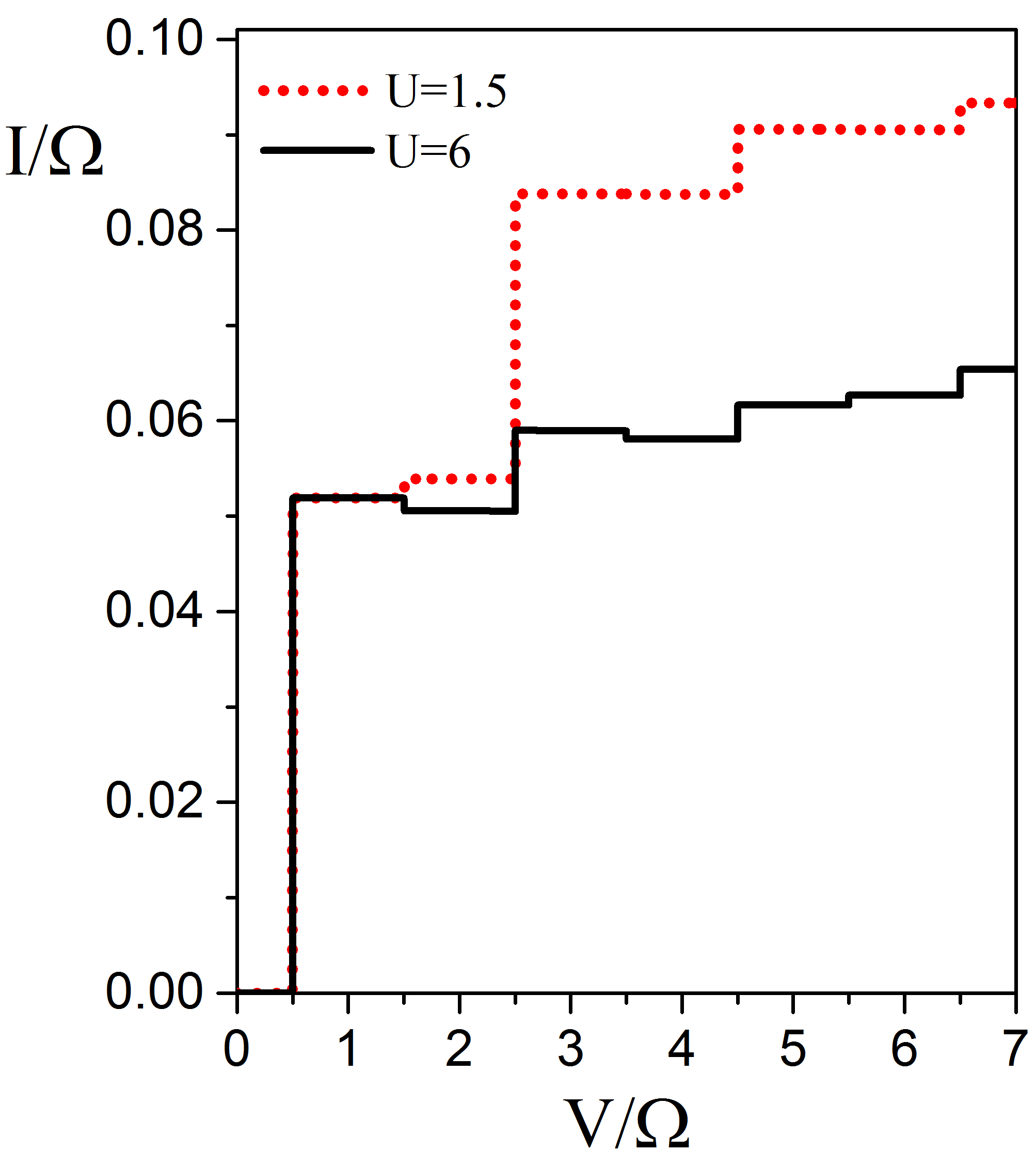}
\caption{\label{fig2} The current as a function of bias voltage, for $ U=1.5 $ and 6, while $ \lambda=0.5 $ and $ \epsilon=-0.25 $. The curves show that for small values of e-e interaction we don't have NDC, however, if the Coulomb repulsion is strong enough, NDC appears at more than one bias voltages.  }
\end{figure}    

In Fig.\ref{fig2} we show the current as a function of bias voltage, for $ U=1.5 $ and 6, while $ \lambda=0.5 $ and $ \epsilon=-0.25 $. From the preceding discussion one can conclude that in order to have NDC, the e-e repulsion should be strong enough. This is confirmed in this plot where for $ U=1.5 $ the system does not show NDC. On the other hand, for strong $ U $s, we can have several bias voltages corresponding to different phonon side-bands, at which NDC appears. As it is shown in the figure, for $ U=6 $, we have NDC at two bias voltages.

\section{Conclusions} \label{con}
In conclusion, we considered an open MJ at zero temperature with both e-e and e-ph interactions. We performed LF transformations and traced out the lead degrees of freedom in the total DM of the system. Using Markov approximation, we obtained the ME that describes the time evolution of our MJ. This ME can be used to obtain the steady state of the system and compute electrical current.

The most important result we obtained was that the system can show NDC as the interplay between e-e and e-ph interactions. For strong enough e-e interactions, this NDC appears when the bias voltage reaches some of phonon side-bands. This phenomena can be understood by noting that at these side-bands the phonons get excited. Even though this opens a new transport channel, because of electron-phonon coupling, tunneling electrons spend more time on the MJ, which increases the Coulomb blockade of opposite spin electrons. It should be noted that in a spin-less model\cite{kosov2017,eskandari2018influence} one can see the increase of electron population on the MJ, however there are no electrons with different spin to be blocked.

From the preceding discussion it is clear that this NDC can only be seen if the e-e interaction is strong enough. However, the e-ph coupling strength should also be in an appropriate range. Very small e-ph interaction will result in negligible current difference which is not useful. On the other hand, for very strong e-ph couplings NDC disappears completely. The reason is at this regime the opened transport channel is more effective and wins the competition. Moreover, the current before the step, $ I_{0} $, is reduced exponentially by increasing e-ph couplings.

%\section*{ACKNOWLEDGMENTS} 

 \appendix
 \section{Lang-Firsov Transformation}   
 In this appendix we drive the Hamiltonian after the LF transformation. This transformation on the operator $ \hat{O} $ is defined as  $\hat{\tilde{O}}= e^{\hat{S}} \hat{O} e^{-\hat{S}} $, where
 \begin{eqnarray}
&&\hat{S}\equiv \sum_{\sigma}\lambda \hat{n}_{\sigma} \left( \hat{b}^{\dag}-\hat{b} \right).
\end{eqnarray}
 We have to obtain the forms of the annihilation and creation operators after this transformation. In order to do that, we define
\begin{eqnarray}
&&\hat{O}^{\eta}=e^{\eta \hat{S}} \hat{O} e^{-\eta \hat{S}},
\end{eqnarray}  
so that $ \hat{O}^{0} $ ($ \hat{O}^{1} $) is the operator $ \hat{O} $ before (after) LF transformation,i.e.,$ \hat{O}^{0}=\hat{O} $ and  $ \hat{O}^{1}=\hat{\tilde{O}} $.

The annihilation and creation operators of the electronic states in the leads are not affected by LF transformation, because they commute with $ \hat{S} $. However, for $ \hat{c}_{\sigma} $ one can show that

\begin{eqnarray}
&&\frac{\partial\hat{c}_{\sigma}^{\eta}}{\partial \eta}\equiv \lambda \left(\hat{b}- \hat{b}^{\dag} \right) \hat{c}_{\sigma}^{\eta}  .
\end{eqnarray}
The formal solution of this differential equation is
\begin{eqnarray}
&&\hat{c}_{\sigma}^{\eta}\equiv\exp [\eta \lambda(\hat{b}-\hat{b}^{\dag})]\hat{c}_{\sigma}.
\end{eqnarray}
 Setting $ \eta=1 $ we obtain the form of the annihilation operator after the LF transformation as
\begin{eqnarray}
&&\hat{\tilde{c}}_{\sigma}=\hat{X}\hat{c}_{\sigma},
\label{ct}
\end{eqnarray}
where $ \hat{X}\equiv\exp [\lambda(\hat{b}-\hat{b}^{\dag})] $. 

Using similar strategy we can show that the form of phonon annihilation operator after LF transformation is
\begin{eqnarray}
&&\hat{\tilde{b}}=\hat{b}-\lambda \sum_{\sigma} \hat{n}_{\sigma}.
\label{bt}
\end{eqnarray}

In order to transform the total Hamiltonian, Eq.\ref{hamt}, we have to transform all of its annihilation and creation operators according to Eqs.\ref{ct} and \ref{bt} and their complex conjugates. This will result in the transformed Hamiltonian, Eq.\ref{lfh}.

\bibliographystyle{model1-num-names}
\bibliography{mp-015.bib}

\end{document}